\documentclass[11pt,epsfig]{article}
\begin{document}
\begin{center}

{ \Large \bf  Synchronization between variable time delayed systems and cryptography }
\\\vskip 5 pt
\end{center}
\vskip 20pt
\begin{center}
{ \it{\large $Dibakar\hskip 2 pt Ghosh^{(a, c)}$\footnote{e-mail: drghosh\_chaos@yahoo.com}, ${Santo\hskip 2 pt Banerjee}^{(b)}$\footnote{e-mail: santo\_banerjee@yahoo.com}and $A. Roy\hskip 2 pt Chowdhury^{(c)}$\footnote{e-mail: asesh\_r@yahoo.com}  }}
     \end{center}
\begin{center}
$^{(a)}$  \it{Department of Mathematics,
                             Dinabandhu Andrews College \\
                              Goria, Calcutta- 700 084, India.}
\end{center}
\begin{center}
$^{(b)}$  \it{Department of Mathematics,
                             JIS College of Engineering,\\
                              Block-A, Phase-3, Kalyani,\\
                                    West Bengal, India.}
\end{center}
\begin{center}
$^{(c)}$  \it{High Energy Physics Division, Department of Physics ,
                            Jadavpur University,\\
                           Calcutta - 700 032, India.}
\end{center}

\begin{center}
\bf{Abstract}
\end{center}
In this letter we consider a prototype model which is described as
an autonomous continuous time delayed differential equation with
just one variable. The chaos has been investigated with variable
delay time and the synchronization phenomenon is examined both
numerically and analytically  using the Krasovskii-Lyapunov functions. We have
applied adaptive coupling law for synchronization,where the coupling
equation also contains delay with modulated time. We also studied
the effect of cryptography for this coupled system and the message
extraction procedure is illustrated with the help of simulated
results.
\section{\bf Introduction:}
Synchronization between two dynamical systems[1] have stimulated a
wide range of research activity [2]. The phenomena of
synchronization in coupled systems have been especially extensively
studied in the context of laser dynamics, electronic circuits,
chemical and biological systems [2]. Application of chaos
synchronization can be found in secure communication, optimization
of nonlinear system performance, modeling brain activity and pattern
recognition phenomena [2]. The most important fields where chaos
could have been practically applied is cryptography. The major
concern in this field is that an encoded message is vulnerable for
extraction by nonlinear dynamic forecasting, when it is hidden by
the signal from the low-dimensional chaotic system. Then it becomes
essential to develop high-dimensional chaotic systems having
multiple number of positive Lyapunov exponents to implement a secure
communication system. In this regard, one time-delayed system could
be of a lot of attention: that is,
\begin{equation}
\dot x = f(x, x_\tau),\hskip 10 pt x_\tau = x(t -\tau)
\end{equation}
For such systems with large constant time-delay $\tau=\tau_0$[3],
some well-established nonlinear
 time-series analysis methods [4, 5, 6] ran into severe problems [7, 8].\\
In this letter we consider a simple one-variable prototype delay system[] containing cubic
nonlinearity. The system has a chaotic behavior for a small constant time delay which was investigated
in []. Here we consider some chaotic properties of this system with variable time delay. We also
investigate the synchronization between two coupled systems with modulated time delay.
The numerical results are supported by analytic calculation of the condition of chaos synchronization
for both scalar and variable delay time with the help of Lyapunov functions.\\
It is known that very simple time-delay systems are able to exhibit
hyperchaos [9]. Consequently it has also been reported that
time-delay systems provide alternative simple and efficient tools
for chaos communication with low detectability []. Chaotic
attractors of time-delay systems can have much higher dimension and
many more positive Lyapunov exponents than the ordinary dynamical
systems.\\
 In this letter, we also show that a message masked by
chaos of a time-delay system with very high dimension and many
positive Lyapunov exponents can be extracted successfully.

\section{Prototype delay model}
The delay differential equation describing the  system are
$$ \dot{x}= \delta x(t-\tau )-\epsilon [x(t-\tau )]^3       \eqno{(2.1)}$$
where $\delta $ and $\epsilon $ are positive system parameters,  $\tau $ is time delay. This system is used as a prototype model to observe self-oscillations in the shipbuilding industry. The chaotic behavior of system (2.1) for constant time delay are studied by A. Ucar []. Here we replace time delay parameter $\tau $ as a function of time instead of constant delay as
$$\tau (t)=\tau _0+asin(\omega t) \eqno{(2.2)}$$ where $\tau _0$ is the zero frequency component, $a$ is the amplitude and $\omega /{2\pi }$ is the frequency of the modulation. The system are in chaotic state for the parameter value $\delta =1,  \epsilon =1,  \tau=1.6 $. The behavior of the system can be change for variable delay time. For fixed value of $\tau _0=1.6$ and $\omega _0=0.8$, we chose the amplitude parameter $'a'$ as bifurcation parameter. The system bifurcation diagram is depicted in figure (1a), in which the system output $x$ stereo-scopically observed from Poincare section and depicted versus the selected range of positive amplitude $'a'$. Figure (1a) shows the trajectory of the system has initially settled down at chaotic state for $a<0.275$. For the range $a\;\varepsilon (0.275, 0.294)$ the system are in periodic states. With increase value of $a$, the system again reach in chaotic state. Further incresing $a(>0.3475)$ the system leads to unbounded solution namely unstable behavior. When the parameter $\delta $ are change then the above bifurcation are changed. The parameter region between $\delta $ and $a$ are depicted in figure (1b). The system trajectory for $a=0.26$ and $\delta =1$ are depicted in figure (1c). In the next section we consider the couple system and calculate analytically the sufficient condition for chaos synchronization.

\section{Coupled system and stability condition}
The couple system can be considered as
$$ \dot{x}= g(x_\tau , \delta , \epsilon )=\delta x(t-\tau )-\epsilon [x(t-\tau )]^3       \eqno{(3.1)} $$
 $$\dot{y}= g(y_\tau , \delta , \epsilon )+k(x-y)=\delta y(t-\tau )-\epsilon [y(t-\tau )]^3+k(x-y)       $$
where $k$ is the coupling strength.
\subsection{scalar delay time}

\par In this section, we study the sufficient condition for synchronization with the help of
 Krasovskii-Lyapunov theory. The desired synchronization manifold is expressed as $x=y$. Let
$\Delta =x-y$ be the synchronization error. Then the dynamics of synchronization error is
$$\dot{\Delta }=-k\Delta +g^{'}(x_\tau , \delta , \epsilon )\Delta _\tau \eqno{(3.2)}  $$
where $g^{'}$ is the derivative of $g$ with respect to time and $\Delta _\tau =\Delta (t-\tau )$. It is obvious that $\Delta =0$ is the trivial solution of equation (3.2) for any value of time delay $\tau .$ To study the stability of synchronization manifold $x=y$, we can use the Krasovskii-Lyapunov functional approach. According to K. Pyragas[PRE, 58, 3067] the sufficient condition for stability of the trivial solution $\Delta =0$ for the time delayed system
$$\dot{\Delta }=-r(t)\Delta +s(t)\Delta _\tau  \eqno{(3.3)}$$
is $r(t)> \mid {s(t)} \mid $. The Krasovskii-Lyapunov functional ( similar to Lyapunov function in the case of ODE) is
$$V(t)=\frac{\Delta ^2}{2}+\mu  \int_{-\tau }^{0}\Delta ^2(t+\theta )d\theta
$$ where $\mu >0$ is an arbitrary positive parameter. The solution $\Delta =0$ is stable if the
derivative of the functional $V(t)$ along the equation (3.3),
$\dot V(t)=-r(t)\Delta ^2+s(t)\Delta \Delta _\tau +\mu \Delta ^2-\mu \Delta ^2_\tau  $  is
negative. $\dot{V}$ will be negative if $4(r-\mu )>s^2$ and $r>\mu >0$. The asymptotic
stability condition for $\Delta =0$ is given for
$$\mu =\frac{\mid s(t)\mid }{2} \;\;\;\;\mbox{and}\;\;\;\; r(t)>\mid s(t)\mid \eqno{(3.4)}$$
For a particular problem, two cases may arise, for the first
case $s$ is constant and $r(t)$ is variable and the second case
$r$ is constant and $s(t)$ is variable. The first case arises in
[Prigas] and in this case $\mu $ is constant.  But for the general
cases the stability condition $\mu =\frac{\mid s(t)\mid }{2}$ is not
always true because $\mu $ is a parameter and $s(t)$ is a variable.
In our case, $r(t)=k$ and $s(t)=g^{'}(x_\tau , \delta , \epsilon )$,
where $s(t)$ is a variable.  So we can not apply the above result.
But the above problem can be removed if we define $\mu $ as a
function of time and the derivative of $\mu $ can be considered in the
expression of $\dot{V}$.
\par Suppose $\mu =\zeta (t)>0$, then the Krasovskii-Lyapunov functional can be taken as [PRE, 75, 037203,2007]
$$V(t)=\frac{1}{2}\Delta ^2(t)+\zeta (t)\int_{-\tau }^{0}\Delta ^2(t+\xi )d\xi  $$
Then $$\dot V(t)=-r(t)\Delta ^2+s(t)\Delta \Delta _\tau +\zeta (t)(\Delta^2 -\Delta ^2_{\tau })+\dot \zeta (t)\int_{-\tau }^{0}\Delta ^2(t+\xi )d\xi$$
If $\dot \zeta (t)\leq 0$ then we have
$$\dot V(t)\leq -[r(t)-s^2(t)/4\zeta (t)-\zeta (t)]\Delta ^2$$
We obtain the stability condition as
$$r(t)-s^2(t)/4\zeta (t)-\zeta (t)>0$$
$$i.e. \;\; r(t)>h(s,\zeta ) \;\;\;\; \mbox{where}\;\;\;\;\; h(s,\zeta )=s^2(t)/4\zeta (t)+\zeta (t)$$
For any function of $s(t)$, $h(s,\zeta )$ is a function of $\zeta (t)$ and has an absolute minimum for $\zeta (t)=\frac{s}{2}$ and $h_{min}(s, \zeta )=\mid s(t)\mid $. Thus $h(s, \zeta )\geq \mid s(t)\mid  $ for any $s$ and $\zeta >0$. The stability condition for synchronization is $r(t)>\mid s(t)\mid $. For system (3.2) the stability condition for synchronization manifold $x=y$ can be written as
$$k>\mid  g^{'}(x_\tau , \delta , \epsilon )\mid \eqno{(3.5)}$$

\subsection{Variable  delay time}
When the time delay of the couple systems is modulated then the sufficient condition for
synchronization are changed. For variable delay time one of the derivative terms in the
expression of $\dot V(t)$ will be included. We define a positive definite Lyapunov functional
of the form
$$V(t)=\frac{1}{2}\Delta ^2(t)+\zeta (t)\int_{-\tau(t) }^{0}\Delta ^2(t+\xi )d\xi  $$
Then  $$\dot V(t)=\Delta  \dot{\Delta}+\dot \zeta (t)\int_{-\tau(t) }^{0}\Delta ^2(t+\xi )d\xi +\zeta (t)[\Delta ^2-\Delta ^2_{\tau }+\Delta ^2_{\tau}\tau ^{'}] $$
if for $\dot \zeta (t)\leq 0$ then $\dot V(t)\leq -\Delta ^2 F(X, \zeta (t))$ where $F(X, \zeta (t))=r(t)-\zeta (t)-s(t)X-[\zeta (t)  \tau^{'}-\zeta (t)]X^2 $ with $X=\frac{\Delta _\tau  }{\Delta } $.
\par In order to show that $\dot V(t)<0$ for all $\Delta  $ and $\Delta _\tau  $  i.e. for all $X$ so it is sufficient to show that $F_{min}>0$.  The absolute minimum of $F$ occurs at $X=\frac{s(t)}{2\zeta (t)(1-\tau ^{'})}$ with $F_{min}=r(t)-\zeta (t)-\frac{s^2(t)}{4\zeta (t)(1-\tau ^{'})}$.
Thus the sufficient condition for synchronization is
$$r(t)>\zeta (t)+\frac{s^2(t)}{4\zeta (t)(1-\tau ^{'})} = \psi (\zeta (t))$$ Again  $\psi (\zeta (t))$ is a function of $\zeta (t)$. The minimum value of $\psi (\zeta (t))$ occurs at $\zeta (t)=\frac{\mid s(t)\mid  }{2\sqrt{1-\tau ^{'}}}$ with $\psi _{min}=\frac{\mid s(t)\mid  }{\sqrt{1-\tau ^{'}}}$.
\par Finally we get the sufficient condition for synchronization is
$$r(t)>\frac{\mid s(t)\mid  }{\sqrt{1-\tau ^{'}}} \;\;\;\; \mbox{with}\;\;\;\;  \zeta (t)=\frac{\mid s(t)\mid }{2\sqrt{1-\tau ^{'}}}  \eqno{(3.6)}$$
Note that in the case of constant delay $\tau ^{'}=\frac{d\tau (t)}{dt}$ vanishes. So the above results are satisfied for constant delay.

\section{Numerical simulation}
In this section, we confirm that the numerical simulation fully can support the analytical results presented above.
For constant coupling parameter and constant delay parameter in the couple system (3.1) we
obtain $r(t)=k$=constant and $\mid s(t)\mid =\mid g^{'}(x_\tau , \delta , \epsilon ) \mid $.
The stability condition (3.5)  takes the form
$k> \mid g^{'}(x_\tau , \delta , \epsilon )\mid $, where the maximum values are taken on
driving
system. The absolute maximum value is
$\frac{2\delta }{\epsilon } \sqrt{\frac{\delta }{3\epsilon  }}$ at $x_\tau =\sqrt{\frac{\delta }{3\epsilon }}$. Thus the analytical condition for chaos synchronization is
$$k>\frac{2\delta }{\epsilon } \sqrt{\frac{\delta }{3\epsilon  }} \eqno{(4.1)}$$ For
the parameter value $\epsilon=1, \tau =1 $ the synchronization region in the
$(k, \delta) $ plane is shown in figure (2a).  At $\delta =1$, for $k=0.2$ and  $k=1.8$ the
synchronization and non-synchronization states are shown in figures (2b) and (2c) respectively.
\par For time delay modulation the analytical condition for synchronization is
$$r(t)>\frac{\mid s(t)\mid  }{\sqrt{1-\tau ^{'}}}\;\;\; \mbox{i.e.}\;\;\;\; k> \frac{2\delta }{\epsilon } \sqrt{\frac{\delta }{3\epsilon  }}\frac{1}{\sqrt{1-a\omega  }} \eqno{(4.2)}$$
For  the set of parameter value $\delta =1, \epsilon =1, a=0.2, \omega =0.8$ the driving system are in chaotic state(fig 1c),
the analytical criterion for chaotic synchronization is $k>0.42$. We take $k=1$, the chaotic
synchronization is shown in figure (3a).
\par When the coupling strength of the couple system (3.1) are varied as
$$ \dot{x}= \delta x(t-\tau )-\epsilon [x(t-\tau )]^3       $$
 $$\dot{y}=\delta y(t-\tau )-\epsilon [y(t-\tau )]^3+k(t)(x-y)   \eqno{(4.3)}     $$
$$\dot{k}(t)=\eta(x-y)^2[k^{*}-k(t-\tau _2)]$$
where $\eta=1.0$ is adaptive gain and $k^{*}=1.9$ is the minimum
coupling strength when coupling are constant and $\tau_2$ is the
coupling delay time between systems $x$ and $y$.  $\tau $ is given
by equation (2.2) and  we take coupling delay $\tau _2$ as modulated
delay time i.e.  $\tau _2=\alpha -\beta  t^2$ where $\alpha $ and
$\beta $ are constants. For $\alpha =2$ and $\beta =1,$ the
variation of $x(t)-y(t)$ and $k(t)$ are shown in figures (3b) and
(3c) respectively.

\section{Cryptography using synchronized system}
It is easy to see from Fig-(3a) that the error vector $w=x-y$ tends
to zero after some period of time $t\ge t_0$ and we have  $x=y$
after $t_0$. We now wish to utilize these two sets of synchronized
chaotic systems for the communication using the cryptographic
encoding. The sender uses system () and the receiver uses system ().
They choose the values of the variables $x$ and $y$ respectively as
the secret keys after some time say t=10.0. To take the values of
the secret keys as integer, they choose [1000x] and [1000y]
respectively. Since they consider the coupled chaotic systems with
time delay, to be more secured they both agree to change the values
of $\tau$ after every eight unit of messages. They both increases
the values of $\tau$ by 0.1 at this interval. The data from $x(t)$
picked up by sender and the corresponding secret keys are shown in
TABLE-1.\\ Let us first describe the procedure how this is
generated. The actual message that a sender wants to send to the
receiver is called plaintext.
 The plaintext and the corresponding ciphertext can be divided into message units. P.G.Vaidya et al showed
  that each unit of message is a single alphabet and they used 26 letter equivalents
   $0,1,\ldots 25$. The corresponding formula for the ciphertext message is
\begin{equation}
c= p+k  mod \hskip 3 pt(26)
 \end{equation}
and decrypted message can be obtained by
\begin{equation}
 p= c-k mod\hskip 3 pt(26)
 \end{equation}

In the letter we have a generalization of the above
procedure.Suppose our message contains both letters as well as
numbers. In references [] the receiver recovered "GOODMORNING" in
state of "GOOD MORNING". An improved version of this method where
the unit of message (it may be a word or a line or a paragraph) is
not only alphabet but also numbers, decimals, space between two
wards or sentences. The corresponding are merely 0,1,\ldots 25 but
more than that.We assign the numeric number 0 to 9 by 0, 1, ....9
respectively. The blank space(gap) between two word is assigned by
10. 11 is the decimal(or full stop between two sentences or any
punctuation marks). The 26 letters from A to Z are assigned from 12
to 37 respectively. The following table will give the clear picture
of the complete assignment.
\begin{center}
\bf{Table-I}
\end{center}
\begin{center}
\begin{tabular}{|c|c|c|c|c|c|c|c|c|c|c|c|c|c|c|c|c|c|c|c|} \hline
Number assigned & 0 & 1& 2& 3& 4& 5& 6 & 7 & 8 & 9 & 10 & 11 & 12& 13  & $\ldots$ &  37 \\
\hline
 Unit Message & 0 & 1 & 2 & 3 & 4 & 5 & 6 & 7 & 8 & 9 & - & . & A & B  & $\ldots$ &  Z \\
\hline
\end{tabular}
\end{center}
The corresponding formula can be written as
$$ c_i= p_i + k_i \hskip 2 pt mod (38) $$
$$ p_i= c_i - k_i \hskip 2 pt mod (38) $$
where $ k_i $ are the secret keys to musk the message. Corresponding
to every message unit we use one and only one message key,which are
randomly generated.For a complete message the secret keys are a
series of numbers $ \{ k_1, k_2, \ldots k_n \} $.Actually the key
$k_j $ hide and secure the message unit $ p_j $.
\par
Let us consider the message which is a sentence that contains numbers also.\\
\begin{center}
 WE \hskip 5 pt HAVE \hskip 5 pt OVER\hskip 5 pt 2700\hskip 5 pt EMPLOYEES.\\
\end{center}
The following table shows the message units and the corresponding
keys.
\begin{center}
\bf{Table-II}
\end{center}
\begin{center}
\begin{tabular}{|c|c|c|c|c|c|c|c|c|c|c|c|c|c|c|c|} \hline
Keys & $k_1$ & $k_2$ & $k_3$ & $k_4$ & $k_5$ & $k_6$ & $k_7$ & $k_8$ & $k_9$ & $k_{10}$ & $k_{11}$ & $k_{12}$ & $k_{13}$ & $k_{14}$ & $k_{15}$ \\
\hline
 Unit Message & w & e & - & h & a & v & e & - & o & v & e & r & - & 2 & 7  \\
\hline
\end{tabular}
\end{center}
\begin{center}
\begin{tabular}{|c|c|c|c|c|c|c|c|c|c|c|c|c|c|} \hline
Keys & $k_{16}$ & $k_{17}$ & $k_{18}$ & $k_{19}$ &  $k_{20} $  & $k_{21}$ & $k_{22}$ & $k_{23}$ & $k_{24}$ &$ k_{25}$ & $k_{26}$ & $k_{27}$ & $k_{28}$  \\
\hline

 Unit Message & 0 & 0 & - & e & m & p & l & o & y & e & e & s & .  \\
\hline
\end{tabular}
\end{center}
Table-3 and Table-4 respectively shows the actual
messages,corresponding plaintext and ciphertext and the message
recovered from the receivers end.
\newpage
\begin{center}
{\bf Table-3} \hskip 10 pt {Ciphertext sent by the sender.}
\end{center}
\begin{center}
\begin{tabular}{|c|c|c|c|c|c|}  \hline
   Time  & Time     &  $x(t)$   &  Keys    &  Plaintext  &   Ciphertext\\
    Lag $\tau_1$ & t  &          &     k    &  p     &
c = p+k mod (38) \\\hline
1.6 & 200.0        & 0.9012..     & 901         & W (34) &  61 \\
      &  201.0        & 0.9265..       & 926      & E (16)   &  23                                        \\
      &   202.0       &  0.9296..      & 929    & --- (10)       & 27                                    \\
       &  203.0         & 0.9327..       & 932      &   H (19)    & 29   \\
      &  204.0        & 0.9515..         & 951   &   A (12)    &   13                                   \\
      &   205.0       & 0.9472..        &  947  &   V (33)      &     68                               \\
     & 206.0          &  0.8412..        & 841 &    E (16)        & 21 \\
      &  207.0        &  0.8281..       & 828  &    --- (10)      &    25                              \\ \hline
 1.8 &   208.0       &  1.4001..       & 1400 &      O (26)      &     42                            \\
      & 209.0         &  1.3874..       & 1387  & V (33)  &  34 \\
      &  210.0        &  1.3574..       & 1357   &   E (16)           &   43                            \\
      &  211.0        & 1.3353..        & 1335    &   R (29)            &     34                         \\
     & 212.0         &  1.3227..        & 1322    &   --- (10)     & 25\\
      & 213.0         & 1.3090..        & 1309      &   2 (2)              &  19                          \\
      &  214.0        & 1.1356..          &  1135       &        7 (7)    &   40                         \\
      &  215.0        & 1.0755..       & 1075      &     0 (0)        &    11                        \\ \hline
  2.00    &  216.0        & 0.47380..       & 473      &   0 (0)          &   17                         \\
      &  217.0        & 0.48044..       &  480     &       --- (10)     &       22                     \\
      &  218.0        & 0.48616..       &   486    &     E (16)   &             31               \\
      &  219.0        & 0.48616..       &   489     &      M (24)       &     57                       \\
      &  220.0        & 0.50019..       &   500    &      P (27)       &       30                     \\
      &  221.0        & 0.50342..       &    503   &      L (23)       &        32                    \\
      &  222.0        & 0.51972..       &     519    &     O (26)       &     51                       \\
      &  223.0       &  0.52531..      &     525    &     Y (36)        &        67                    \\ \hline
 2.20 &  224.0        & 0.96425..       &    964    &     E (16)        &       23                     \\
      & 225.0         & 0.95570..       &     955    &     E (16)        &         21                   \\
      & 226.0        &  0.93889..      &      938   &      S (30)      &          43                  \\
      & 227.0         & 0.91450..       &     914   &     . (11)        &
      12
      \\ \hline

\end{tabular}

\end{center}
\newpage
\begin{center}
{\bf Table-4} \hskip 10 pt {Plaintext recovered by the receiver.}
\end{center}
\begin{center}
\begin{tabular}{|c|c|c|c|c|c|}  \hline
   Time  & Time     &  $y(t)$   &  Keys    &  Ciphertext  &   plaintext\\
    Lag $\tau_1$ & t  &          &     k    &  p     &
p = c-k mod (38) \\\hline
1.6 & 200.0        & 0.9012..     & 901         & 61 &  34 (W) \\
      &  201.0        & 0.9265..       & 926      & 23   &  16 (E)                                        \\
      &   202.0       &  0.9296..      & 929    & 27       & 10 (---)                                    \\
       &  203.0         & 0.9327..       & 932      &   29    & 19 (H)   \\
      &  204.0        & 0.9515..         & 951   &   13    &   12 (A)                                   \\
      &   205.0       & 0.9472..        &  947  &   68      &     33 (V)                               \\
     & 206.0          &  0.8412..        & 841 &    21        & 16 (E) \\
      &  207.0        &  0.8281..       & 828  &    25      &   10 (---)                               \\ \hline
 1.8 &   208.0       &  1.4001..       & 1400 &      42      &  26 (O)                               \\
      & 209.0         &  1.3874..       & 1387  & 34  &   33 (V)\\
      &  210.0        &  1.3574..       & 1357   &   43           &    16 (E)                           \\
      &  211.0        & 1.3353..        & 1335    &   34            &       29 (R)                       \\
     & 212.0         &  1.3227..        & 1322    &   25     & 10 (---)\\
      & 213.0         & 1.3090..        & 1309      &  19              &       2 (2)                     \\
      &  214.0        & 1.1356..          &  1135       &        40    &    7 (7)                        \\
      &  215.0        & 1.0755..       & 1075      &     11        &      0 (0)                      \\ \hline
  2.00    &  216.0        & 0.47380..       & 473      &   17          &    0 (0)                        \\
      &  217.0        & 0.48044..       &  480     &       22     &          10 (---)                  \\
      &  218.0        & 0.48616..       &   486    &     31   &              16 (E)             \\
      &  219.0        & 0.48616..       &   489     &      57       &       24 (M)                    \\
      &  220.0        & 0.50019..       &   500    &      30       &          27 (P)                  \\
      &  221.0        & 0.50342..       &    503   &      32       &          23 (L)                  \\
      &  222.0        & 0.51972..       &     519    &     51       &       26 (O)                      \\
      &  223.0       &  0.52531..      &     525    &     67        &           36 (Y)                  \\ \hline
 2.20 &  224.0        & 0.96425..       &    964    &     23        &         16 (E)                   \\
      & 225.0         & 0.95570..       &     955    &     21        &            16 (E)                \\
      & 226.0        &  0.93889..      &      938   &     43    &          30 (S)                  \\
      & 227.0         & 0.91450..       &     914   &     12        &
      11 (.)
      \\ \hline

\end{tabular}

\end{center}

\section{Conclusion}
In this communication, we have studied the chaotic behavior of
prototype delayed model with variable time delay. The detailed
parameter region and bifurcation diagram are obtained when the time
delay of the system are modulated. The synchronization between two
unidirectionally coupled chaotic systems with fixed and variable
time delay are analyzed. we have also obtained the sufficient
condition for chaos synchronization manifold by K-L function and it
was verified by numerical simulation. In the last section  we
considered the case of variable coupling in the modulated delay
situation and have shown how it can be utilized in the process of
communication using cryptography.
\section{Acknowledgement}
Santo Banerjee is thankful to Prof.R.Ramaswamy, Department of
physics, Jawaharlal University, India for his valuable and important
successions.

\section{References}
{1}. A. Ucar, "A prototype model for chaos studies" International Journal of Engineering Science {40} {(2002)} {251-258}.\\
{[1]} L. M. Pecora and T. L. Carroll, Phys. Rev. Lett. 64, 821 (1990); E.Ott, C.Grebogi and
J.A.Yorke, Phys.Rev.Lett. 64, 1196 (1990).\\
{[2]} CHAOS, Special issue on chaos synchronization 7,N4 (1997) edited by W.L.Ditto and
K.Showalter; G.Chen and X.Dong, From Chaos to Order.Methodologies, Perspectives
and Applications (World Scientific, Singapore,1998); Handbook of Chaos Control, Ed.
H.G.Schuster (Wiley-VCH, Weinheim,1999).\\
{[3]}. K. Pyragas, Phys. Rev. E 58, 3067 (1998); R. He and P.G.
Vaidya, Phys. Rev. E 59, 4048 (1999); L. Yaowen, G, Gguangming, Z.
Hong, and W. Yinghai,
Phys. Rev. E 62, 7898 (2000).\\
{[4]}. P. Grassberger, and I. Procaccia, Physica D 9, 189 (1983).\\
{[5]}. J. P. Eckmann, and D. Ruelle, Rev. Mod. Phys. 57, 617 (1985).\\
{[6]}. M. B. Kennel, R. Brown, H. D. I. Abarbanel, Phys. Rev. A 45, 3403 (1992).\\
{[7]}. Th. Meyer, N. H. Packard, in Nonlinear Modeling and
Forecasting, edited by M. Casdagli,
and S. Eubank (Addison-Wesley, Redwood City, CA, 1992).\\
{[8]}. R. Hegger, H. Kantz, and Olbrich, in Proceedings of the
Workshop on "Nonlinear Tech- niques in Physiological Time Series
Analysis," edited by H. Kantz, J. Kurths, and G.
Mayer-Kress (Springer, Berlin, 1997).\\
{[9]} J.D. Farmer,Physica D 4,366 (1982). K.M. Short and A.T.
Parker, Phys. Rev. E 58, 1159 (1998).\\
{[a]} N. N. Krasovskii, Stability of Motion ( Stanford University Press, Stanford, 1963).\\
{[b]} K. Pyragas, Phys. Rev. E. 58, 3067 (1998).\\

\section{Caption of figures}
Figure (1a) : Bifurcation diagram with respect to the amplitude parameter $a$ when other parameters are $\delta =1, \epsilon =1, \tau =1.6, \omega =0.8$.\\
Figure (1b): Parameter region between $\delta $ and $a$ where gray region is period one, light black region period two, deep black region are period three and white represent chaotic region.\\
Figure (1c) : Phase space in the plane $(x, \dot x(t))$ when
$a=0.26$. \\
Figure (2a): Parameter region between coupling parameter $k$ and
system parameter $\delta$. Shaded portion represent synchronization
region. \\
Figure (2b): Time variation  of the relative error  $w=x-y$, which
indicates the system are
not synchronized.\\
Figure (2c): Time series of $x-y$ which indicates the system are
synchronized.\\
Figure (3a): Time variation of the relative error $w$ where the
delay parameter $\tau$ is a function of time.\\
Figure (3b): Time evolution of $w$ corresponding to the equation
(4.3).\\
 Figure (3c): Time variation of  the coupling parameter $k$.

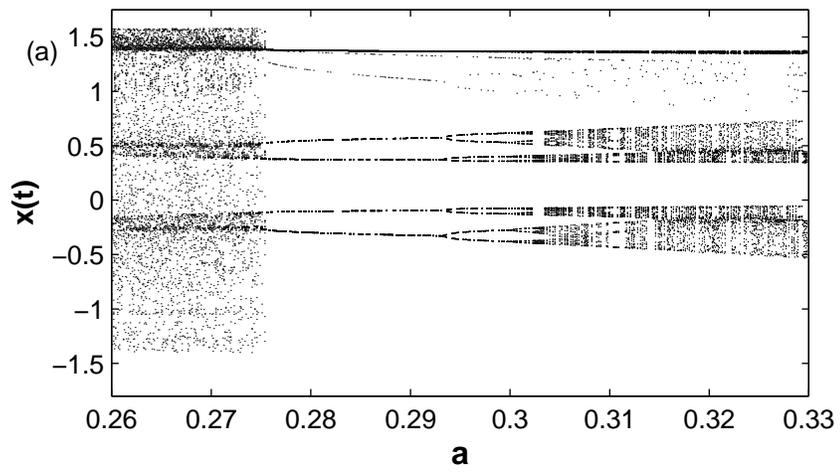
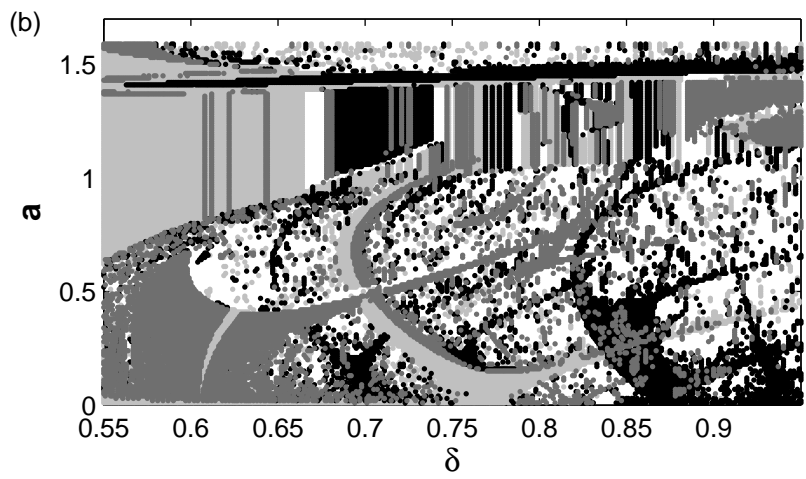
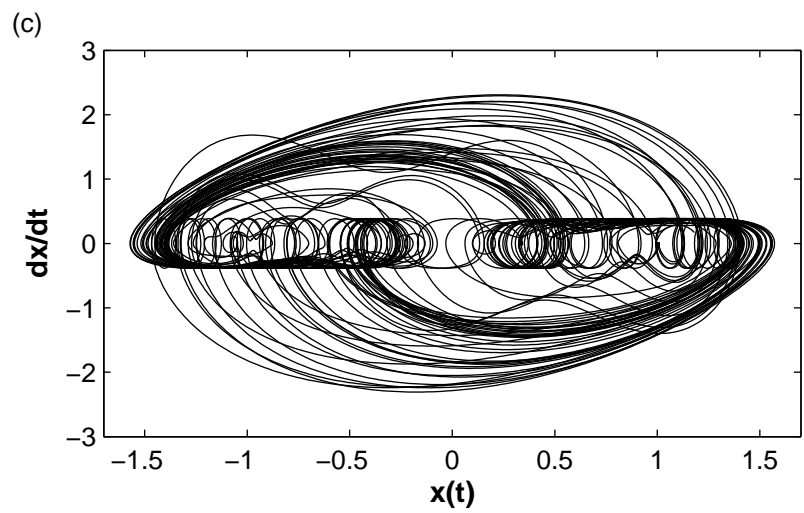
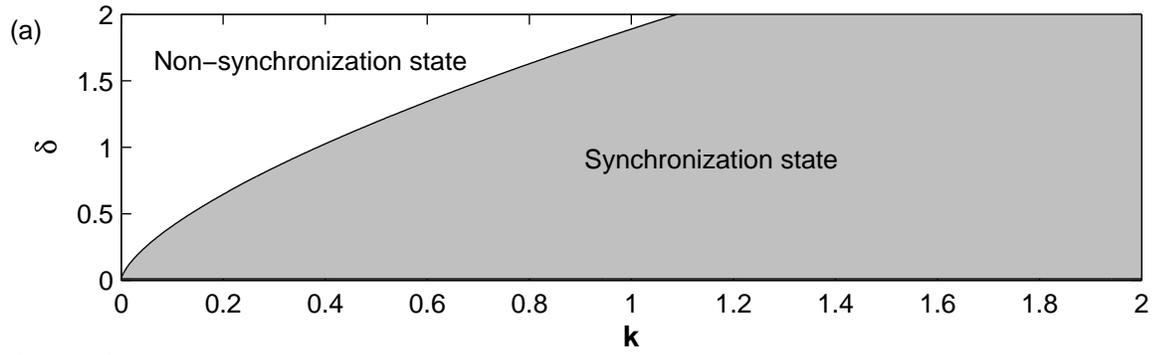
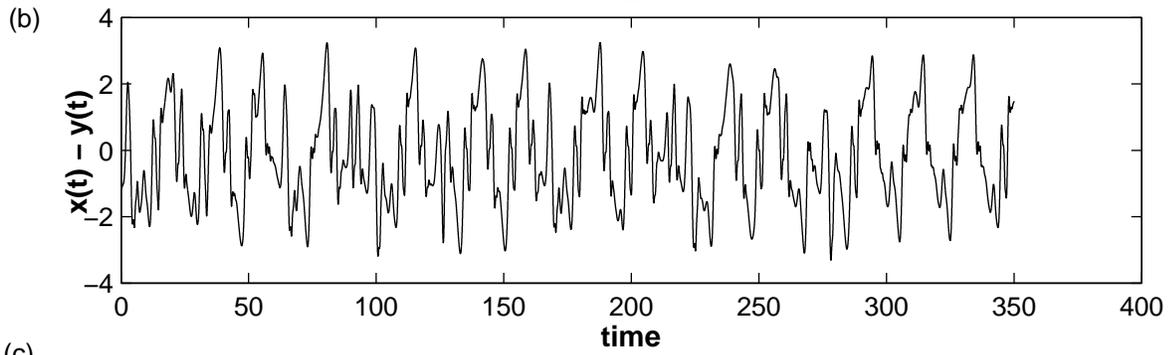
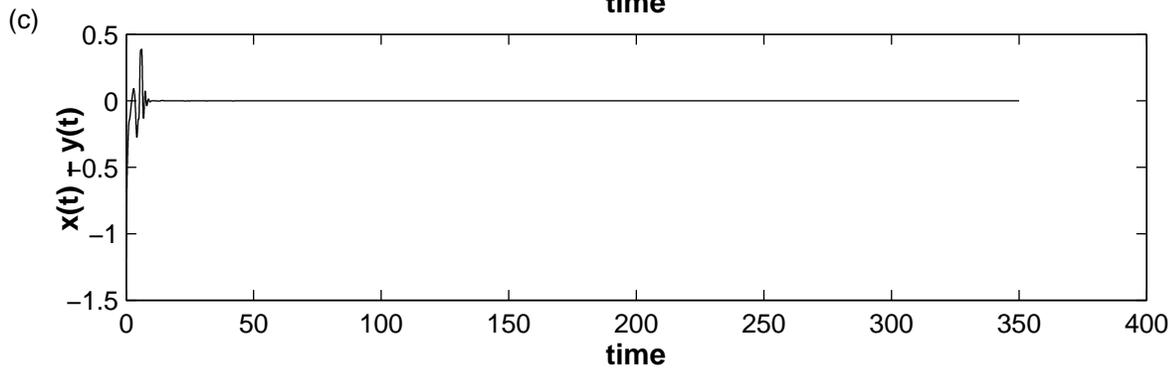
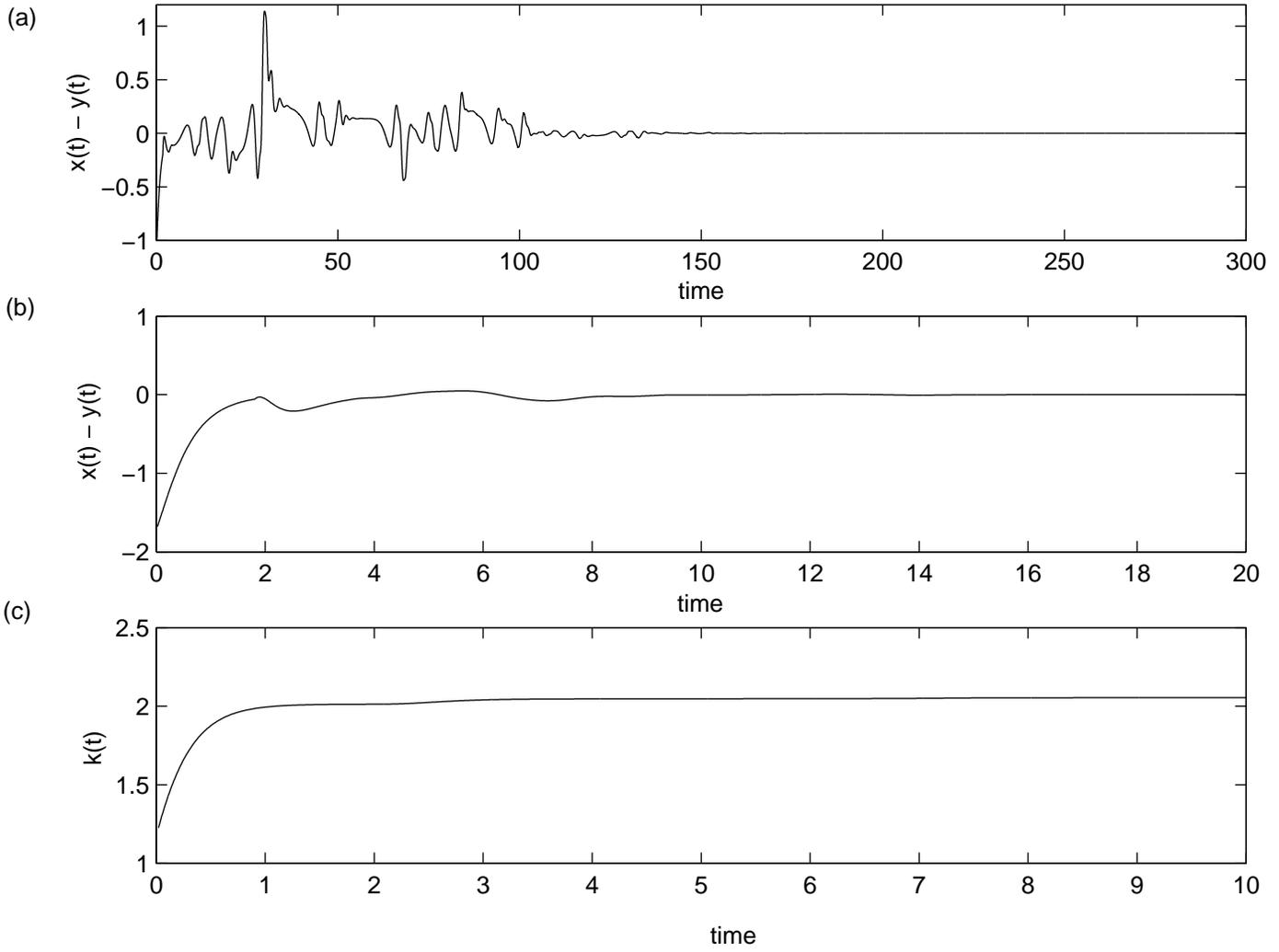
\end{document}